\documentclass[letterpaper,twocolumn,english,prb,showpacs,amsmath,amssymb]{revtex4-1}
\usepackage[latin9]{inputenc}
\usepackage{amsmath}
\usepackage{amssymb}
\usepackage{graphicx}
\usepackage{esint}

\makeatletter

 \@ifundefined{textcolor}{}
 {%
   \definecolor{BLACK}{gray}{0}
   \definecolor{WHITE}{gray}{1}
   \definecolor{RED}{rgb}{1,0,0}
   \definecolor{GREEN}{rgb}{0,1,0}
   \definecolor{BLUE}{rgb}{0,0,1}
   \definecolor{CYAN}{cmyk}{1,0,0,0}
   \definecolor{MAGENTA}{cmyk}{0,1,0,0}
   \definecolor{YELLOW}{cmyk}{0,0,1,0}
 }

\usepackage{epsfig}\usepackage{subfigure}\usepackage{dcolumn}\usepackage{stmaryrd}\usepackage{mathrsfs}\usepackage{pifont}
\usepackage{amsthm}\usepackage{bm}\usepackage{latexsym}\usepackage{amsfonts}

\newcommand{\beq}{\begin{equation}}
\newcommand{\eeq}{\end{equation}}
\newcommand{\beqa}{\begin{eqnarray}}
\newcommand{\eeqa}{\end{eqnarray}}

\usepackage{babel}

\makeatother

\usepackage{babel}
\begin{document}

\title{Toward high-fidelity coherent electron spin transport in a GaAs double quantum dot}
\author{Xinyu Zhao}
\author{Xuedong Hu}
\email{xhu@buffalo.edu}
\selectlanguage{english}%
\affiliation{Department of Physics, University at Buffalo, SUNY, Buffalo, New York 14260-1500, USA}
\begin{abstract}
In this paper, we investigate how to achieve high-fidelity electron spin transport in a GaAs double quantum dot.  Our study examines spin transport from multiple perspectives.  We first study how a double dot potential may affect/accelerate spin relaxation.  We calculate spin relaxation rate in a wide range of experimental parameters and focus on the occurrence of spin hot spots. A safe parameter regime is identified in order to avoid these spin hot spots. We also study the non-adiabatic transitions in the Landau-Zener process of sweeping the interdot detuning, and propose a scheme to take advantage of possible Landau-Zener-St\"{u}kelburg interference to achieve high-fidelity spin transport at a higher speed. Finally, we calculate the double-dot correction on the effective $g$-factor for the tunneling electron, and estimate the resulting phase error between different spin states. Our results should provide a useful guidance for future experiments on coherent electron spin transport.
\end{abstract}

\pacs{73.63-b, 72.25.Rb, 03.67.Hk.}

\maketitle

\section{Introduction}

In universal quantum computing, quantum information inevitably needs to be transferred over finite distances on chip or between chips. For spin qubits in semiconductor nanostructures,\cite{Loss1998, Oosterkamp1998, Hayashi2003PRL, Petta2004PRL, Stavrou2005PRB, Hanson2007, Bluhm2011, Veldhorst2014, Muhonen2014} there are a variety of ways such long-distance communication can be achieved.\cite{Barnes2000, Skinner2003, Taylor2005, FriesenPRL2007, TrifPRB2008, HuPRB2012}  One particularly straightforward way is to coherently move the electrons themselves between quantum dots. Indeed, coherently transporting electrons between quantum confined states, with their spin states intact, could be a critical component of a wide range of future quantum coherent devices that utilize the electron spins.

There are two major approaches to achieve coherent transport of spin qubits, one using surface acoustic waves,\cite{Barnes2000, Hermelin2011, McNeil2011, Sanada2011, Yamamoto2012, Huang2013, Bertrand2016, Bertrand2016a, Zhao2016, Hermelin2017} the other by tuning the electric potentials on a series of surface gates.\cite{Lu2003, Skinner2003, Greentree2004, Taylor2005, Wang2013, Baart2016, Fujita2017npjQI}  We have studied the former in the past,\cite{Huang2013, Zhao2016} and will in this paper focus on the latter, which is an integral part of a concerted experimental effort towards making larger arrays of quantum dots \cite{Zajac2016, Baart2016, Fujita2017npjQI}. Indeed, the importance of coherent spin transport goes well beyond quantum information transfer.  Other important quantum operations, such as error correction and spin readout, also involve electron tunneling between quantum dots.\cite{VanHouten1996, Kim2014, OGorman2016, Pica2016, Baart2016, Cao2016}  In the broader context of semiconductor heterostructures, an investigation of transport properties between quantum dots and nanowires is also an important element in the search and control of possible Majorana fermion excitations.\cite{Weymann2017, Albrecht2017PRL}

Practically, quantum tunneling of an electron is usually driven by tuning the bias voltage between neighboring quantum dots. During such a process, several factors could change the spin state of the electron and reduce the fidelity of spin transfer. For example, spin relaxation due to spin-orbit interaction (SOI) \cite{Bychkov1984,Dresselhaus1955} and phonon emission could be modified by the double-dot confinement as opposed to a single-dot confinement.\cite{Stano2005}  The degeneracy near zero bias causes an energy level anti-crossing, so that a time-dependent Hamiltonian from sweeping the electric field with a finite speed could cause non-adiabatic transitions, which also reduce the fidelity of the electron spin transfer. Furthermore, the SOI together with the confinement potential causes corrections to the eigen-energies, leading to a small modification of the effective $g$-factor, which could be significant if a superposed spin state is being transferred.

In this work, we study how to achieve high-fidelity spin and charge transfer through electron tunneling in a double dot. In particular, we examine how interdot tunneling affects spin relaxation, and identify the parameter range where spin hot spots can be avoided. In the regime where spin relaxation effect is minimized, we study how spin transfer fidelity can be maximized in the Landau-Zener process of sweeping the interdot detuning potential, and how pulse shaping can help increase the transfer fidelity. We also propose a scheme to achieve high-speed electron transport through Landau-Zener interference. Such a scheme can also be used to measure the tunnel barrier between the two dots. Last but not least, we study the effective $g$-factor with a correction caused by SOI and the double dot potential, and point out that missing this correction can cause a significant error in the tracking of the phase difference between spin up and down states.

The paper is organized as follows. In Sec.~\ref{sec:Model}, we describe the double quantum dot model we consider and explain the protocol for interdot electron transport. In Sec.~\ref{sub:relaxation}, we investigate spin relaxation during the tunneling process. In Sec.~\ref{sec:LZ}, we study the Landau-Zener processes in the spin transport as we sweep the interdot detuning. Particularly, we study the interference between two adjacent Landau-Zener processes in Sec.~\ref{sec:LZS}, and explore the possibility of using interference to increase fidelity. Last but not least, we investigate corrections on the effective $g$-factor in Sec.~\ref{sec:Effective g}. Finally, we discuss our results and draw some conclusions in Sec.~\ref{sec:Conclusion}.

\section{\label{sec:Model}Model of a double quantum dot}

\begin{figure}
\includegraphics[width = 1\columnwidth]{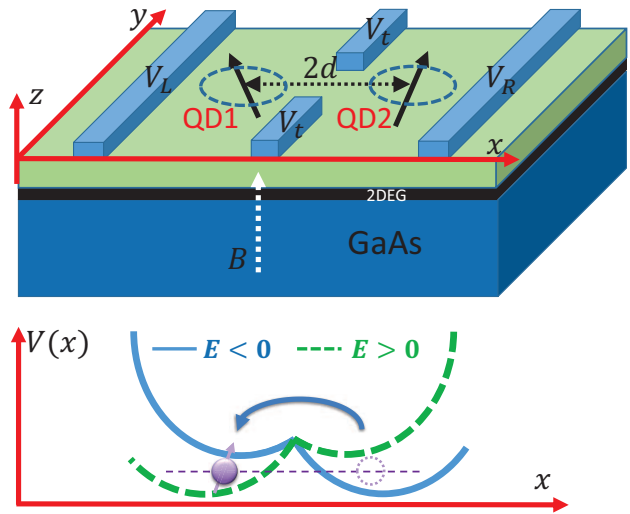}

\caption{\label{fig:sketch}(color online) Sketch of our protocol for electron transport in a double quantum dot. The two-dimensional DQD resides at the interface of GaAs and the barrier material, with the growth-direction confinement much stronger than the in-plane confinement. The regions ``QD1'' and ``QD2'' label the two dots. Surface gates $V_{L}$ and $V_{R}$ can be used to adjust the detuning between the two dots, while $V_{t}$ can be used to tune the tunnel coupling strength. }
\end{figure}

As discussed in the Introduction, in this paper we study electron spin transport that is enabled by tuning the applied voltages on the metallic surface gates. While a dense array of gates together with optimized programming of voltages can probably achieve relatively smooth motion of a quantum dot potential, here we focus on a much simpler protocol. Assuming the existence of a double quantum dot (DQD) potential, as illustrated in Fig.~\ref{fig:sketch}, changing the interdot detuning via an applied electric field shifts the ground orbital state from one dot to the other, thereby achieving electron transport. In such a process, the only time-dependent variable is the electric field applied across the DQD, tunable by one or two surface gates.

The system we consider is a two-dimensional GaAs DQD with an electric field applied along the interdot axis. The confinement along the growth direction is much stronger so that we do not consider any excitation in that direction. The system Hamiltonian is thus given by
\begin{equation}
H=T+V_{0}+H_{E}+H_{Z}+H_{SO}+H_{hf}\,,\label{eq:Htot}
\end{equation}
where
\begin{eqnarray}
T & = & \frac{\hbar^{2} \mathbf{\pi}^{2}}{2m^{*}}\,,\label{eq:kineticE}\\
V_{0}(x,y) & = & \frac{1}{2}m\omega_{0}^{2}[(|x|-d)^{2}+y^{2})]\,,\label{eq:V0}\\
H_{E} & = & eEx\,,\\
H_{Z} & = & \frac{1}{2}g\mu_{B}B\sigma_{z}\,,\\
H_{SO} & = & \frac{\alpha_{BR}}{\hbar}(\sigma_{x}\pi_{y}-\sigma_{y}\pi_{x})+\frac{\alpha_{D}}{\hbar}(\sigma_{y}\pi_{y}-\sigma_{x}\pi_{x})\,,\,\\
H_{hf} & = & \frac{1}{2}g\mu_{B} \mathbf{B}_{nuc}\cdot \mathbf{\sigma} \,.\label{eq:Hhf}
\end{eqnarray}
Here $\mathbf{\pi} = \mathbf{p}+e\mathbf{A}/\hbar$ is the kinetic momentum, $m^{*}$ the effective mass of the electron, $e$ the electron charge, and
$\mathbf{A}=B(-y/2,x/2,0)$ the vector potential of the applied magnetic field. The external magnetic field is applied along the $z$-direction
(growth direction), which introduces a Zeeman splitting given by $H_{Z}$.  The double quantum dot confinement potential is modeled by a double
harmonic $V_{0}$,\cite{Tarucha1996, Kouwenhoven1997, Stano2005} where $d$ gives the half interdot distance. In this simple model, varying the interdot distance also changes the tunnel barrier between the two dots. The interdot detuning $V_d=2eEd$ is controlled by an electric field via $H_{E}$, which in practice can be tuned by voltages applied on gates $V_{L}$ and $V_{R}$, as shown in Fig.~\ref{fig:sketch}.  $V_{0}$ and $H_{E}$ together gives the total electric potential $V=V_{0}+H_{E}$, which is schematically plotted in the bottom panel of Fig.~\ref{fig:sketch} in two cases: $E>0$ (blue solid line) and $E<0$ (green dashed line). The electron transport is achieved by tuning the electric field $E$. In other words in our protocol $E=E(t)$. We assume the change of the electric field is sufficiently slow as compared to the single-electron excitation energy, so that the electron undergoes an adiabatic transfer from the right dot to the left dot.

Lastly, $H_{SO}$ and $H_{hf}$ describe two major mechanisms of spin mixing. $H_{SO}$ is the spin-orbit coupling, where $\alpha_{D}$ and $\alpha_{BR}$ are the strength of Dresselhaus and Bychkov-Rashba SOI, respectively.\cite{Dresselhaus1955, Bychkov1984, DeSousa2003, Stano2005}  In the following calculations, we use $\alpha_{D}=4.5\;\rm{meV}\cdot\AA$ and $\alpha_{BR}=3.3\;\rm{meV}\cdot\AA$, as in Ref.~\onlinecite{Stano2006}. $H_{hf}$ is the hyperfine interaction between the electron and the environmental nuclear spins. In our calculation we take the lowest order mean-field approximation, where the effect of the nuclear spins is modeled as an extra magnetic field $\mathbf{B}_{nuc}$, the Overhauser field. Under normal experimental conditions, the Overhauser field is in an arbitrary direction and is position-dependent. Generally the $z$-component of $\mathbf{B}_{nuc}$ causes a small modification of the Zeeman energy, and the $x-y$ components make spin-flip transitions possible.

To obtain spin transfer fidelity in our protocol, we solve the time evolution of the electron state governed by the time-dependent Hamiltonian.  To account for non-adiabatic effects, we go beyond the lowest-energy orbital states, making our calculation quite complex and inevitably numerical. Instead of solving the time-and-space-dependent Schr\"{o}dinger equation directly, we first solve for the instantaneous eigenstates $\psi_{m}(t)$ and eigenenergies $\epsilon_{m}(t)$ by numerically diagonalizing the Hamiltonian $H(t)$ at an electric field $E(t)$ for a series of points in time, then solve the time-evolution problem by expanding on the basis of the instantaneous eigenstates $\psi(x,y,t)=\sum_{m}C_{m}(t)\psi_{m}(x,y,t)$, so that the Schr\"{o}dinger equation becomes
\begin{align}
 & i\hbar\frac{\partial}{\partial t}C_{m}(t)=C_{m}(t)\epsilon_{m}(t)\nonumber \\
 & -i\hbar\sum_{n}C_{n}(t)\int dxdy \ \psi_{m}^{*}(x,y,t) \frac{\partial}{\partial t} \psi_{n}(x,y,t).
\end{align}
This approach becomes particularly transparent as we approach the adiabatic limit, when the electron would evolve following the instantaneous
eigenstates.

\begin{figure}
\includegraphics[width=1\columnwidth]{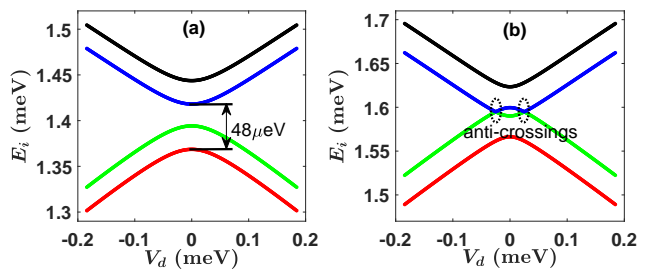}

\caption{\label{fig:EngLevel}(color online) Typical energy diagrams for a DQD.  The interdot detuning is given by $V_{d}=2eEd$, with $d=46.4nm$. The magnetic field is $B=1{\rm T}$ for (a) and $B=1.3\mbox{T}$ for (b). In panel (a), from top to bottom, the black, blue, green, and red lines represent the energy levels of the states $|e,\downarrow\rangle$, $|e,\uparrow\rangle$, $|g,\downarrow\rangle$, and $|g,\uparrow\rangle$. For $B=1.3\mbox{T}$, when Zeeman energy is larger than the tunnel coupling, the energy of $|e,\uparrow\rangle$ is smaller than the energy of $|g,\downarrow\rangle$ near $V_{d}=0$.}
\end{figure}

In Fig.~\ref{fig:EngLevel}~(a) we plot a typical low-energy diagram of the DQD. When spin-orbit mixing is negligible, from top to bottom, the four curves represent the energy levels of the states $|e,\downarrow\rangle$, $|e,\uparrow\rangle$, $|g,\downarrow\rangle$, and $|g,\uparrow\rangle$, where $|g\rangle$ is ground orbital state and $|e\rangle$ is the first excited orbital state, $|\uparrow\rangle$ and $|\downarrow\rangle$ indicate the spin states. Essentially each orbital state splits into two parallel spin branches. When $E\ll0$ ($V_{d}\ll0$), the ground orbital state $|g\rangle$ is approximately the lowest-energy Fock-Darwin state located in the right dot $|\psi_{R}\rangle\propto\exp\{ [-(x+d)^{2} - y^{2}]/2a^{2}\}$,
and the excited state $|e\rangle$ is approximately the ground Fock-Darwin state located in the left dot $|\psi_{L} \rangle \propto \exp \{ [-(x-d)^{2} - y^{2}]/2a^{2}\}$, where $a=\left( \hbar/m^{*} \sqrt{\omega_{0}^{2} + \omega_{c}^{2}/4} \right)^{1/2}$ is the effective confinement length, with $\omega_{c}=eB/m^{*}$.  When $E \gg 0$ ($V_{d} \gg 0$), the ground state and excited states are switched, and the left dot Fock-Darwin state $|\psi_{L}\rangle$ becomes the ground state. Near the zero detuning $V_d=0$, $|g\rangle$ and $|e\rangle$ are mixtures of $|\psi_{L} \rangle$ and $|\psi_{R} \rangle$, and an anti-crossing forms with an energy gap $2t_{E}$. This makes our protocol essentially a Landau-Zener process, which will be analyzed in detail in the following sections.

With the correction of Zeeman energy, the spin-up excited state $|e,\uparrow\rangle$ could have equal or even lower energy than the state $|g,\downarrow\rangle$ near $V_d=0$ when the magnetic field is above the threshold given by the tunnel coupling. An example is given in Fig.~\ref{fig:EngLevel}~(b) for a relatively large $B$ field. Near zero detuning, two anti-crossings between different spin states are formed. Through SOI, spin states are mixed near the two anti-crossings in Fig.~\ref{fig:EngLevel}~(b), which allow transitions between eigenstates of the far-detuned limit. We will discuss the consequences of these anti-crossings in the next section.

\section{\label{sec:Transport}Spin transfer fidelity in a double quantum
dot}

The objective of our protocol is to transfer the complete spin information from one dot to the other at the fastest rate. This transfer entails
the transfer of both the carrier itself, i.e. the electron, and the spin state. Obviously, multiple factors could affect the fidelity of this spin transfer. In this Section we will study these factors one by one.

First of all, decoherence could destroy the spin state. In GaAs, both the hyperfine interaction and the SOI could cause spin relaxation.  For hyperfine interaction, we estimate the worst spin relaxation caused by the transverse field. For SOI, we examine how it could be enhanced by the nearby excited states when the two dots are nearly symmetric, especially the strong relaxation at the anti-crossings of different spin states.

Second, with a finite speed for transportation that we would like to push as fast as we can, the electron could be excited to higher orbital and/or spin states, which reduces spin transfer fidelity. We will study these unwanted transitions and propose possible schemes to enhance the transport fidelity, by either weakening these transitions or using interferences to suppress their net effect.

Third, the SOI introduces corrections in the single-electron energy levels, and such a correction can change the dynamical phase between different spin states when a superposition state is transported.
We will show in the last subsection that the correction on the effective $g$-factor can cause a notable phase error during the transport, so that one has to keep track of it in order to maintain the correct superposition.

\subsection{\label{sub:relaxation}spin relaxation}

Spin relaxation in quantum dots generally involve two major interactions: SOI or hyperfine interaction to mix the spin states, and electron-phonon
interaction to facilitate transitions between states with different energies. In this subsection, we calculate the rate of spin relaxation caused by both SOI and hyperfine interaction, and show that relaxation can be neglected if experimental parameters are chosen properly.

We first investigate the relaxation caused by the hyperfine interaction.\cite{Khaetskii2000PRB,Erlingsson2001PRB,Johnson2005Nature}
As discussed in the previous Section, we treat the hyperfine interaction within the mean field approximation, so that the nuclear spin effects are fully represented by the Overhauser field $\mathbf{B}_{nuc}$.  The longitudinal part of $\mathbf{B}_{nuc}$, or the $z$ component, causes inhomogeneous broadening for the electron spin because nuclear spins are quasi-static on the time scale of tunneling (nanoseconds).\cite{Petta2005Science}  Electron motion allows the electron spin to sample more nuclear spins, therefore reducing their dephasing effect via motional narrowing, as discussed in Ref.~\onlinecite{Huang2013}.  The transverse part of the Overhauser field, on the other hand,

slightly tilts the quantization axis for the electron spin $H_{spin} = \frac{1}{2} g\mu_{B} (B\sigma_{z} + B_{nuc, xy} \sigma_{x})$, and causes a spin in an original eigenstate to precess.  Within the spin precession cycle, the minimum fidelity is (when the spin has the largest deviation from the original eigenstate)
\begin{equation}
F_{min}=\frac{B^{2}}{B^{2}+B_{nuc,xy}^{2}}\,.\label{eq:Fmin}
\end{equation}
In a typical GaAs quantum dot, $B_{nuc}$ (or $B_{nuc, xy}$) is estimated to be $2-6\;\mbox{mT}$,\cite{Johnson2005Nature} while the external field
$B$ is typically much larger, at least a fraction of a Tesla.  Equation (\ref{eq:Fmin}) would give a fidelity of 0.9999 if $B$ is about 100 times larger than $B_{nuc, xy}$.  Therefore, the fidelity loss caused by the transverse part of the Overhauser field can always be neglected under normal experimental conditions.

In the rest of this subsection we focus on spin relaxation caused by SOI. Spin flip due to spin-orbit coupling and phonon emission is usually the most important spin relaxation mechanism for a quantum dot confined electron spin in GaAs \cite{Khaetskii2001PRB, Stano2006, Amasha2008, Dugaev2009, Raith2012, Srinivasa2013}.  When the electron is being transported with a constant velocity, Doppler effect causes modifications to the spin relaxation rate and angular distribution of the emitted phonons \cite{Dugaev2009,Huang2013,Zhao2016}.  However, in the present case of an electron moving in a double dot, the speed of motion is quite slow and the Doppler shift negligible.  Our focus is thus more on how interdot coupling may modify the spin-phonon coupling and spin relaxation under quasi-static condition, and the transition rates we calculate are between instantaneous eigenstates.

Spin mixing is already included in our calculation of the instantaneous eigenstates when we diagonalize Hamiltonian Eq.~(\ref{eq:Htot}) that contains SOI.  For electron-phonon interaction we consider both deformation potential and piezoelectric interaction between the confined electron and the
acoustic phonon environment. The interaction Hamiltonians are
\begin{eqnarray}
H_{df} & = & \Sigma_{e}\sum_{k}\sqrt{\frac{\hbar k}{2\rho Vc_{1}}}e^{i \mathbf{k} \cdot \mathbf{r}}(b_{k,1}+b_{-k,1}), \\
H_{pz} & = & -ih_{14}\sum_{k,\lambda}\sqrt{\frac{\hbar}{2\rho Vc_{\lambda}k}}M_{\lambda}e^{ik\cdot r}(b_{k,\lambda}+b_{-k,\lambda}^{\dagger})\,.
\end{eqnarray}
Here $\lambda=1,2,3$ indicates phonon polarization (1 for the longitudinal mode, while 2 and 3 for the two transverse modes), $\mathbf{k} = (k_{x},k_{y},k_{z})$ is the phonon wave vector, $\Sigma_{e}=7{\rm eV}$ is the GaAs deformation potential, $h_{14}=1.4\times10^{9}{\rm eV/m}$ is the piezoelectric constant, $\rho=5.3\times10^{3}kg/m^{3}$ is the mass density, $c_{1}=5.3 \times 10^{3}m/s$ and $c_{2} = c_{3} = 2.5 \times 10^{3}m/s$
are the speeds of sound for longitudinal and transverse phonons in bulk GaAs, and $b_{k,\lambda}$ and $b_{k,\lambda}^{\dagger}$ are the annihilation and creation operators for phonons in mode $\lambda$ and with wave vector $\mathbf{k}$. The piezoelectric interaction matrix element is $M_{\lambda} = 2(k_{x} k_{y} e_{z}^{\lambda} + k_{z} k_{x} e_{y}^{\lambda} + k_{y}k_{z} e_{x}^{\lambda})$, where $e_{x}^{\lambda}$, $e_{y}^{\lambda}$, $e_{z}^{\lambda}$ are the components of the unit polarization vectors.

Given the electron-phonon interaction Hamiltonian and the electron eigenstates, the relaxation rate between two eigenstates can be computed by Fermi's golden rule as
\begin{equation}
\Gamma_{df}=\left[\bar{n}+1\right]\frac{\sigma_{e}^{2}\epsilon_{fi}^{2}}{8\pi^{2}\rho c_{1}^{4}\hbar^{3}}\int d^{2}k\;|\langle\psi_{f}|e^{ik\cdot r}|\psi_{i}\rangle|^{2}/k_{z}^{1}
\end{equation}
\begin{equation}
\Gamma_{pz}=\left[\bar{n}+1\right]\sum_{\lambda}\frac{(h_{14})^{2}}{8\pi^{2}\hbar\rho c_{\lambda}^{2}}\int d^{2}k\;|M_{\lambda}|^{2}|\langle\psi_{f}|e^{ik\cdot r}|\psi_{i}\rangle|^{2}/k_{z}^{\lambda}
\end{equation}
where $\epsilon_{fi}$ is the energy difference between the initial state ($|\psi_{i}\rangle$) and the final state ($|\psi_{f}\rangle$), and $\bar{n}$ is the thermal occupation number of the phonon state at the energy $\epsilon_{fi}$, which is approximately zero for most Zeeman splitting at the dilution fridge temperature.

\begin{figure}
\includegraphics[width=1\columnwidth]{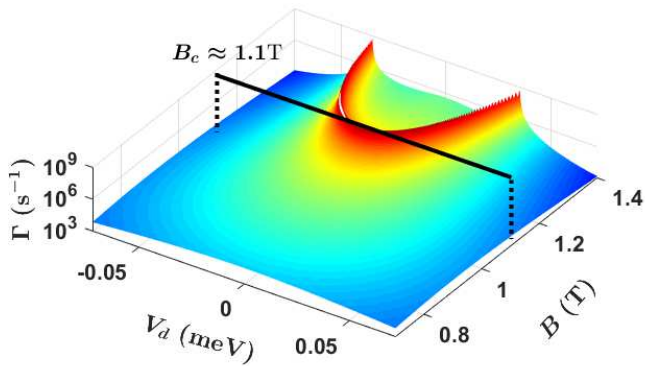}

\caption{\label{fig:RRvsEB}(color online) Spin relaxation rate as a function of the applied electric and magnetic field. Interdot detuning $V_d=2eEd$ with $d=48\mbox{nm}$, and single dot single-particle excitation energy is $\hbar \omega_{0} \approx 1.1$ meV. }
\end{figure}

In general, electron spin relaxation rate depends on the applied electric and magnetic fields.  The electric field changes the composition of the states, therefore modifying the matrix elements within the integrands of the relaxation rates above; while magnetic field changes the Zeeman splitting directly, therefore affecting the range of the integrals.  In Fig.~\ref{fig:RRvsEB} we plot the overall relaxation rate $\Gamma = \Gamma_{df} + \Gamma_{pz}$ as a function of both $E$- and $B$-field.  The most prominent features are the sharp peaks for the relaxation rate, which are called spin hot spots \cite{Stano2006,Hu2011,Raith2012,Srinivasa2013}.  The relaxation rate at these peaks are in the order of GHz, on par with a normal charge qubit.  These hot spots are produced by the SOI-induced anti-crossing between states $|g,\downarrow\rangle$ and $|e,\uparrow\rangle$.  At these anti-crossings spin is not a good quantum number, so that the relaxation rate is determined by the charge relaxation matrix element between $|g\rangle$ and $|e\rangle$.

The electric- and magnetic-field dependence of the hot spots are quite straightforward.  In a low magnetic field, tunnel splitting is the dominant energy at zero detuning: $2t_{E} \gg E_{Z} = g\mu B$, the energy difference between $|g,\downarrow\rangle$ and $|e,\uparrow\rangle$ is too large to allow any significant mixing, therefore no hot spots. As magnetic field increases toward $B_{c}$ that satisfies $g\mu_{B}B_{c} = 2t_{E}$, the energies of state $|g,\downarrow\rangle$ and $|e,\uparrow\rangle$ become close to each other at zero detuning, and a SOI-induced anti-crossing starts to form
between the two states.  Consequently a single spin hot spot appears at $B=B_{c}$ and $V_d=0$.  In a higher magnetic field, the energy of $|g,\downarrow \rangle$ is larger than that of $|e,\uparrow \rangle$ at zero detuning ($V_d=0$), so that two anti-crossings form symmetrically on either side of the zero detuning point. The resulting maximum mixture at the anti-crossings produce the two relaxation peaks \cite{Srinivasa2013} in Fig.~\ref{fig:RRvsEB} for a given magnetic field $B > B_{c}$ (One appears at $V_d<0$, the other symmetrically at $V_d>0$).

Incidentally, the fact that a spin hot spot appears at $B \geqslant B_{c}$ can be used to detect the tunneling matrix element $t_{E}$. A similar method has been used to detect valley splitting in a Si quantum dot.\cite{Yang2013}

\begin{figure}
\includegraphics[width=1\columnwidth]{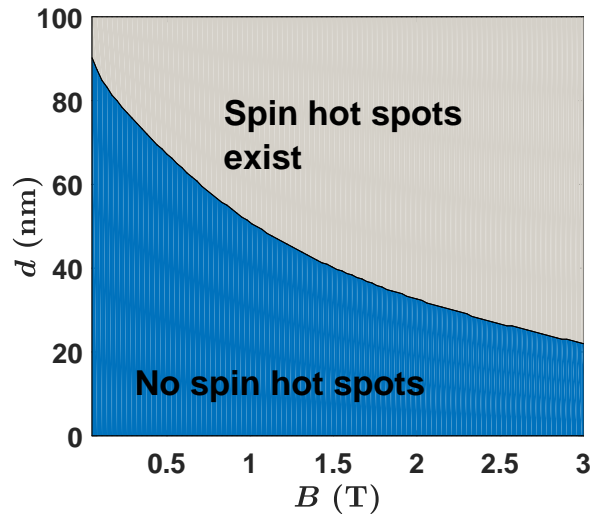}

\caption{\label{fig:HotRegion}(color online) Spin relaxation hot spots in the parameter space of half interdot distance $d$ and magnetic field $B$, two important adjustable parameters in experiment. Other parameters are the same as in Fig.~\ref{fig:RRvsEB}.}
\end{figure}

With our double-harmonic model of a DQD potential, the tunnel splitting $t_{E}$ is determined by the interdot distance $d$: the larger the $d$ is, the higher and wider the tunnel barrier, the lower the $t_{E}$.  We could thus define a safe region in the parameter space expanded by $d$ and $B$, where spin hot spots are absent. In Fig.~\ref{fig:HotRegion} this safe region is the bottom-left blue region, where spin relaxation rate is in the order of $\Gamma \approx 10^{3}{\rm Hz}$. In the upper-right gray region, hot spots would appear at a certain electric field. Near and at the hot spots, the relaxation rate rapidly increases to the level of $\Gamma \approx 10^{9}{\rm Hz}$, similar to the relaxation rate of a charge qubit. As mentioned above, the boundary between the two regions is roughly given by the condition $t_{E} = E_{Z} = \frac{1}{2}g\mu_{B}B$.

Our results demonstrate that spin relaxation rate is generally quite low in a DQD, in the order of $\Gamma\lesssim10^{3}{\rm Hz}$ at lower magnetic field, if we can avoid spin hot spots. Under such conditions spin relaxation would not be an important issue for high fidelity spin transport since tunneling generally happens at the nanosecond time scale. Furthermore, even if an experiment is performed in the ``unsafe'' region, we note that the hot spots in Fig.~\ref{fig:RRvsEB} are quite sharp, so that fast spin relaxation only appear within a small range of $E$-field. Thus we should be able to keep spin relaxation error small as long as the time we sweep through a hot spot is sufficiently brief.  In short, in most cases spin relaxation does not cause any significant issue to high-fidelity spin transport, especially if the experimental parameters are tuned to the safe region as suggested in Fig.~\ref{fig:HotRegion}.

\subsection{\label{sec:LZ}Landau-Zener transitions}

In quantum mechanics, Landau-Zener transitions occur when a time-dependent system Hamiltonian is swept through a level anti-crossing. In this
subsection, we study how the fidelity of our spin transport protocol may be affected by Landau-Zener transitions.\cite{rubbmark1981PR, shevchenko2010PhysRep, petta2010Science, studenikin2012PRL, korkusinski2017PRL}

In a Landau-Zener (LZ) transition, the diabatic transition probability, i.e. the probability that the quantum state does not follow the adiabatic
path, is given by \cite{rubbmark1981PR}
\begin{equation}
P_{D}=\exp\left(-\frac{2\pi\Delta E_{nm}^{2}/\hbar}{d|E_{n}-E_{m}|/dt}\right)\,.\label{eq:PD}
\end{equation}
Here $\Delta E_{nm}=(E_{n}-E_{m})/2$ (at $min\{E_{n}-E_{m}\}$) is half of the energy gap at the anti-crossing point, and \mbox{$d|E_{n}-E_{m}|/dt$}
is the time derivative of the gap between the two anti-crossing levels $n$ and $m$ as the Hamiltonian is swept through the anti-crossing.

As illustrated in Fig.~\ref{fig:EngLevel}~(a), in our spin transport protocol, the electric field is swept from negative to positive, in the middle of which an orbital-level anti-crossing is formed. If the electric field is increased too fast, unwanted diabiatic transitions will lead to finite probabilities of excitation into excited final states. For example, if the electron is initially in the ground state of the right dot, one possible final excited state is when the electron remains in the right dot ground state and fails to tunnel.  Furthermore, Figure \ref{fig:EngLevel}~(b) shows that at higher magnetic fields, two SOI-induced anti-crossings are also present, giving rise to additional possible diabatic transitions that may or may not be desirable.

In this subsection we focus on possible LZ transitions between orbital states as shown in Fig.~\ref{fig:EngLevel}~(a). We choose the energy gap between the ground $|g,\uparrow\rangle$ (or $|g,\downarrow\rangle$) and excited orbital state $|e,\uparrow\rangle$ (or $|e,\downarrow\rangle$) to be about 10 GHz.  More precisely, $2\Delta E_{31} \approx 2\Delta E_{42} \approx 48\mu\mbox{eV}$.  Taking a linearly increasing electric field $E(t) = E_{0} \frac{t}{T}$ ($-T\leqslant t\leqslant T$), $d|E_{n}-E_{m}|/dt$ is inversely proportional to the total operation time $2T$, so that the probability of diabatic transition $P_{D}$ is an exponentially decaying function of the total operation time $2T$.

\begin{figure}
\includegraphics[width=1\columnwidth]{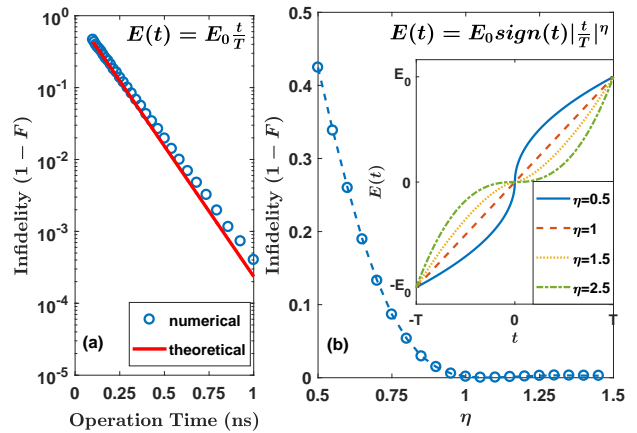}

\caption{\label{fig:fvslambda}(color online) (a) Infidelity of spin transport as a function of the total operation time for a linear pulse $E(t) = E_{0} \frac{t}{T}$. (b) Transport infidelity for different power-law time dependence of the electric field pulse. The pulse shape is determined by the power $\eta$ as in $E(t) = E_{0} \mbox{sign}(t) |t/T|^{\eta}$.  For panel (b), the total operation time is fixed at $2T = 0.8$ ns. For both panels (a) and (b) $E_{0}=4000$ V/m, $d=46.4$ nm, $B=1$ T, and $\hbar\omega_{0}=1.1$ meV.}
\end{figure}

In Fig.~\ref{fig:fvslambda}~(a), we compare the numerical results of the fidelity defined as $F = |\langle \psi_{i} | \psi_{f} \rangle|^{2}$ (plotted as infidelity $1-F$) and the theoretical prediction from Eq.~(\ref{eq:PD}).  Notice that the Landau-Zener formula (\ref{eq:PD}) agrees quite well with the numerical simulation of the dynamics, even though Eq.~(\ref{eq:PD}) is derived for a simple two-level model \cite{rubbmark1981PR, shevchenko2010PhysRep}, while our double dot model is complicated by factors such as higher orbital states and corrections from SOI.  Clearly, the corrections from all the complexities are relatively small and the dynamics of the double dot can be roughly modeled as a two-level (orbital) system. One simple observation we can make here is that in order to achieve high fidelity transport, the time duration of the field-sweep should be sufficiently long to avoid unwanted transitions.  Specifically, for the orbital LZ transition considered here, a total operation time longer than $0.5$ ns for a detuning change of $0.74$~meV could ensure a 0.99 fidelity.

A linearly varying electric field is far from optimal in ensuring adiabatic electron tunneling between the DQD.  As indicated in Eq.~(\ref{eq:PD}), one can modify the shape of the detuning voltage pulse in order to keep the system in the ground state.  Keeping the total evolution time as a constant, one can design a pulse that changes more slowly near the minimum gap and more quickly away from zero detuning.  As an illustration we numerically study several pulses described by $E(t)=E_{0}{\rm sign}(t)\left|\frac{t}{T}\right|^{\eta}$, where time $t$ changes from $-T$ to $T$, and the function ${\rm sign}(t)=1$ for $t\geqslant0$, ${\rm sign}(t)=-1$ for $t<0$. The larger the power $\eta$ is, the slower the $E$-field changes near $E=0$, as shown in the inset of Fig.~\ref{fig:fvslambda}~(b). The resulting infidelity $1-F$ of the evolution is presented in Fig.~\ref{fig:fvslambda}~(b). As expected, a larger $\eta$ gives rise to a slower evolution near the anti-crossing, which leads to a higher fidelity. Notice that here we have simply chosen a few power-law functions as an illustration, without any attempt at optimization. There are certainly better pulse shapes to avoid or enhance a transition. One can also design alternative techniques to modify the system evolution. For example, adding extra control pulses can also help remove non-adiabatic contributions, and achieve ``shortcuts to adiabaticity''.\cite{Chen2010,Ban2012}

\subsection{\label{sec:LZS}Landau-Zener-St\"{u}ckelburg interferences}

Landau-Zener-St\"{u}ckelburg (LZS) interference could occur when a system Hamiltonian is swept through multiple anti-crossings.\cite{rubbmark1981PR, shevchenko2010PhysRep, petta2010Science, studenikin2012PRL, korkusinski2017PRL}  In this subsection we explore how LZS interference may help spin transfer at higher magnetic fields.

\begin{figure}
\includegraphics[width=1\columnwidth]{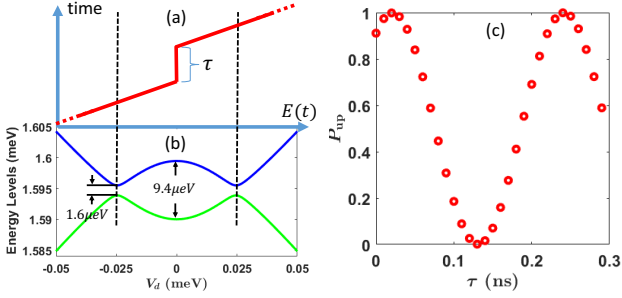}

\caption{\label{fig:pup}(color online) (a) Schematic diagram of pulse shape for tuning LZS interference. A plateau is inserted in an otherwise linearly increasing electric field. The duration of the plateau is given by parameter $\tau$. (b) Energy diagram near zero detuning. The full energy diagram is given in Fig.~\ref{fig:EngLevel}~(b) with two other levels. The actual electric field increases from $-1500$ V/m ($V_{d} = 2eEd \approx -0.14$ meV) to $1500$ V/m ($V_{d} \approx 0.14$ meV). (c) Spin-up probability ($P_{up}$) in the final state after an extra waiting time $\tau$ is inserted at the point $V_d=0$ in the electric field pulse. The other parameters are chosen as $B=1.3$ T, $d=46.4$ nm, total operation time without counting $\tau$ is $2T-\tau = 20$ ns.}
\end{figure}

As shown in Fig.~\ref{fig:EngLevel}~(b), when $B > B_{c}$, states $|g,\downarrow\rangle$ and $|e,\uparrow\rangle$ would cross at certain detunings, and spin would mix because of SOI. The resulting anti-crossings mean that unwanted LZ transitions between the two spin states could occur as we sweep the system Hamiltonian through either one of them. The relatively weak SOI in GaAs means that 
the gap for these anti-crossings are much smaller than the orbital anti-crossing gap. 
Thus a complete adiabatic evolution requires a much slower sweeping speed.  Conversely,
a fast passage through these anti-crossings would keep the electron spin unchanged,
which is desirable for spin transport. For each of the SOI-induced anti-crossings in Fig.~\ref{fig:pup}~(b), the minimum energy gap between $E_{2}$ and $E_{3}$ is estimated at $2\Delta E_{32}=1.6{\rm \mu eV}$.  The gap is computed by taking the SOI parameters as $\alpha_{D}=4.5\;\rm{meV}\cdot\AA$ and $\alpha_{BR}=3.3 \;\rm{meV}\cdot\AA$ (see Ref.~\onlinecite{Stano2006} and experimental references therein). According to Eq.~(\ref{eq:PD}), a diabatic transition
probability $P_{D}\approx\frac{1}{2}$ is possible if we sweep the electric field from $-1500\mbox{V/m}$ to $1500\mbox{V/m}$ (corresponding to $V_{d} = 2eEd$ changing from $-0.14\mbox{meV}$ to $0.14\mbox{meV}$) in about $20$ ns. A numerical simulation confirms this estimate (not shown in the figure).

The presence of two LZ processes near each other opens the possibility of guiding the electron state towards a desired outcome using LZS interference.  The energy diagram here is similar to the structure of a two-paths interferometer that is widely used in quantum optics \cite{Stehlik2012, Rancic2016, Gallego-Marcos2016, Osika2017, schoenfield2017NatComm}.  The two energy levels here are similar to the two arms of an interferometer, while the dynamical phase between the energy levels is analogous to the phase difference between two optical paths. These analogies indicate that we should be able to control the output state by manipulating the phase difference between the two energy levels.

One way to control the interference is to add a waiting period $\tau$ at $E=0$ to the original linearly-increasing pulse $E(t)=E_{0}\frac{t}{T}$ in Fig.~\ref{fig:pup}~(a).  Consider an initial state in the second eigenstate $|2\rangle$. After the first LZ process, the state becomes a superposition $|2\rangle \rightarrow \sqrt{P_{A}} |2\rangle + \sqrt{P_{D}} |3\rangle$, where $P_{A}$ and $P_{D}$ are the adiabatic and diabatic transition probabilities. After the evolution between the two Landau-Zener processes, the state becomes $\sqrt{P_{A}}|2\rangle + e^{i(\phi+\delta\phi)} \sqrt{P_{D}} |3\rangle$, where $\phi$ is the normal dynamical phase accumulated between the two anti-crossings, while $\delta\phi$ is the extra phase that can be controlled by the duration of the wait at $E=0$. Eventually, after the second LZ process, the state becomes (unnormalized)
\begin{equation}
|\psi_{out}\rangle=(P_{A}+P_{D}e^{i(\phi+\delta\phi)})|2\rangle+\sqrt{P_{A}P_{D}}(1+e^{i(\phi+\delta\phi)})|3\rangle.
\end{equation}
In order to obtain an output state $|2\rangle$, the extra phase needs to satisfy $1+e^{i(\phi+\delta\phi)}=0$. One can also obtain $|3\rangle$ by properly choosing a different extra phase.  As shown in Fig.~\ref{fig:pup}~(c), the output state indeed undergoes the LZS interference and oscillates between spin up and down determined by the extra phase between the two Landau-Zener processes. As shown in Fig.~\ref{fig:pup}, the probability of obtaining a spin-up state can reach $1$ if an extra phase is properly chosen.  Alternatively, this interference may also be employed to achieve controlled spin flip.

The interference pattern in Fig.~\ref{fig:pup}~(c) contains useful information about the DQD.  Specifically, the period of the spin state oscillation is roughly $\Delta \tau = 0.22{\rm ns}$, which indicates that the energy splitting between the 2nd and the 3rd level at zero detuning should be $(E_{3}-E_{2})|_{E=0}=9.4\mu{\rm eV}$, since the additional phase difference is given by $\delta\phi=(E_{3}-E_{2})\Delta\tau/\hbar$.  From the numerical result shown in Fig.~\ref{fig:pup}~(b), the zero-detuning energy gap is indeed close to the value predicted from the interference pattern. An accurate measurement of this energy splitting thus gives further information on the tunnel barrier between the two dots.

\subsection{\label{sec:Effective g}Effective $g$-factor}

\begin{figure}
\includegraphics[width=1\columnwidth]{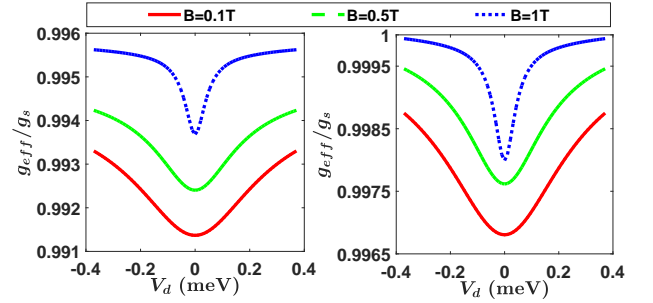}

\caption{\label{fig:geff}The DQD effective $g$-factor $g_{eff}$ as a function of the applied electric and magnetic field.  The half interdot distance is again $d=46.4nm$, and $\hbar \omega_0 = 1.1$ meV. The left panel gives the ratio of $g_{eff}$ over the bulk value, while the right panel is the ratio of $g_eff$ over the effective $g$-factor $g_{s}$ of a single dot.}
\end{figure}

In this subsection we investigate a correction on the $g$-factor of the electron spin during the transport, and show that this correction leads to a non-negligible modification to the dynamical phase of the spin.

Typically, for a spin qubit the orbital degree of freedom is frozen, which means that the electron should always be in the ground orbital state. During spin transport the electron should also remain in the instantaneous ground orbital states.  We can therefore define an effective $g$-factor based
on the energy difference between states $|g,\uparrow\rangle$ and $|g,\downarrow\rangle$ as
\begin{equation}
g_{eff}=(\epsilon_{g,\downarrow}-\epsilon_{g,\uparrow})/\mu_{B}B \,.
\end{equation}
Without SOI, orbital motion and spin evolve in their own Hilbert sub-space separately, so that the energy difference above is exactly the Zeeman energy $g\mu_{B}B$, and $g_{eff}$ is equal to the bulk value $g$. However, with SOI in a QD or DQD, both $\epsilon_{g,\downarrow}$ and $\epsilon_{g,\uparrow}$ are modified slightly, and the effective $g$-factor also deviate slightly from the bulk value.

In Fig.~\ref{fig:geff} we plot $g_{eff}$ as a function of the interdot detuning $V_{d} = 2eEd$. In the left panel $g_{eff}$ is normalized against the bulk $g$-factor, while in the right panel it is normalized against the single-dot effective $g$-factor $g_{s}$. Figure~\ref{fig:geff} clearly shows that $g_{eff}$ depends on both the applied magnetic field and the electric field/interdot detuning, and the corrections are the largest at the zero-detuning point. Quantitatively, the correction on the $g$-factor is smaller than $1\%$ in the whole parameter space.  Considering that the initial state in a transport experiment is always prepared in the lowest-energy orbital state in a single dot, the results from right panel should be more directly relevant in evaluating the effects of a double dot potential.

The enhancement of the correction on the electron $g$-factor near zero detuning is due to the more even distribution of the orbital wave function across the two dots. Take the spin-orbit Hamiltonian $H_{SO}$ as a perturbation to $H=T+V_{0}+H_{Z}+H_{E}$, the second-order perturbation gives a correction on the $i^{th}$ energy level $\delta E_{i} \sim -(\alpha_{D}^{2}-\alpha_{BR}^{2}) [1-\langle i |\sigma_{z} (x\pi_{y} - y\pi_{x}|i\rangle]$.  Mathematically, a wave function distributed more evenly in the two dots has a larger mean value of $\langle i|\sigma_{z} (x\pi_{y} - y\pi_{x}) |i\rangle$, leading to a stronger correction on the $g$-factor.

Any change in the electron $g$-factor would modify the dynamical phase between its two spin orientations. While such a modification does not matter to a spin eigenstate, it could be significant for a superposed state. For example, suppose we are to transport a superposed state $|g,\uparrow\rangle + |g,\downarrow\rangle$, a phase factor would appear in the final state $|g, \uparrow \rangle + e^{i\Phi}|g, \downarrow \rangle$, where
\begin{equation}
\Phi=\int_{0}^{T}\frac{1}{\hbar}g_{eff}(\tau)\mu_{B}Bd\tau.
\end{equation}
Here, $g_{eff}(\tau)$ is a time-dependent function because $E(\tau)$ changes with time and $g_{eff}$ depends on $E$. This phase $\Phi$ has to be tracked accurately in order to maintain high fidelity of the spin state. Clearly, if one was to use the bulk $|g|=0.44$ or the single-dot $g$-factor to calculate $\Phi$, the phase information becomes inexact. We can define a phase error as $\Phi_{error} = \int_{0}^{T} \frac{1}{\hbar} \left[ g_{eff} (\tau) - g \right] \mu_{B} B d\tau$, where $g=g_{{\rm bulk}}$ or $g=g_{s}$.  Numerical results show that the accumulated phase error could be non-negligible even though the correction on $g_{eff}$ is always smaller than $1\%$, since $\Phi_{error}$ is an integration over time.
For example, the time average of $g_{eff}$ at $B=1T$ is about $\bar{g}_{eff} \approx 0.999g_{s}$ (compared to $g_{s}$) or $\bar{g}_{eff} \approx 0.995g$ (compared to $g$). If we use the bulk value $g$ to estimate the phase, a 10 ns operation time will cause a phase error of $0.62\pi$.
If we use the single dot value $g_{s}$ to estimate the phase, the error will reach $0.12\pi$ when the operation time is 10 ns. Therefore,
in the calculation of dynamical phase, corrections on $g$ factor must be taken into consideration.

Both a modified $g$-factor and a random longitudinal Overhauser field cause corrections to the dynamical phase of a superposed spin state, and need to be addressed when transporting a coherent spin state.  In a 1 T applied field their effects are also similar in order: a 0.1\% correction on the $g$-factor is equivalent to a 1 mT change in the magnetic field, while the magnitude of Overhauser field in a typical GaAs quantum dot is about 2 mT. On the other hand, the two physical mechanisms are qualitatively different from the perspective of quantum coherence. The effect of $g$-factor is systematic and is completely determined by the SOI coupling strength and the double dot potential, so that it can be calculated \textit{a priori}. The longitudinal Overhauser field, on the other hand, is random and can only be determined through direct measurement, even though the slow dynamics of Overhauser field allows its lowest-order effect to be eliminated via spin echo.

\section{\label{sec:Conclusion}Discussion and Conclusion}

In conclusion, we have investigated how to maintain high fidelity when transporting an electron spin qubit in a GaAs DQD.
In particular, we have studied spin relaxation caused by the SOI, and show that spin hot spots are present in high magnetic fields. We identify the reason behind spin hot spots as SOI-induced level mixing between different spin states. We give a safe region in parameter space to avoid spin hot spots using the guideline $g\mu_B B < t_E$.  In the regime where spin relaxation effect is minimized, we demonstrate how spin transfer fidelity can be maximized in the Landau-Zener process of sweeping the interdot detuning potential, and how pulse shaping can help increase the transfer fidelity. We also propose a scheme to achieve high-speed electron transport through Landau-Zener-St\"{u}ckelburg interference. Such a scheme can also be used to measure the tunnel barrier between the two dots. Last but not least, we study the effective $g$-factor with a correction caused by SOI and the double dot potential.  We point out that while this correction in $g$-factor is always under 1\%, missing the correction can cause a significant error in the tracking of the phase difference between spin up and down states.

Spin transport in other semiconductor materials should have qualitatively similar behaviors, though quantitatively the differences from GaAs could be significant. For example, in InSb, which has much larger SOI \cite{Winkler2003}, the energy correction would be more significant, which means larger anti-crossing gaps and larger corrections on the $g$-factor. Spin transport could also be important for testing and manipulating Marjorana fermion excitations in nanowires \cite{Weymann2017,Albrecht2017PRL}.

Silicon quantum dots \cite{Yang2013,Huang2014a} present another interesting challenge to spin transport. The nearly degenerate valley states in the conduction band could introduce significant additional complexities into spin transport. Specifically, spin-valley mixing can cause a new type of spin relaxation \cite{Yang2013}, while the extra valley degree of freedom can produce additional anti-crossings and interference between these anti-crossings \cite{Kim2014}. Such new features and challenges will be investigated elsewhere.

\begin{acknowledgments}
We acknowledge financial support by US ARO through grants W911NF1210609 and W911NF1710257.
\end{acknowledgments}

\bibliographystyle{apsrev4-1}
\bibliography{PaperGaAsBIB}

\end{document}